
\input harvmac
\Title{EFI-93-21}
{\vbox {\centerline{The Light Cone in String Theory}
}}

\centerline{\it Emil Martinec\footnote{*}{Supported
in part by Dept. of Energy grant DEFG02-90ER-40560.  }
}
\smallskip
\centerline{Enrico Fermi Inst. and Dept. of Physics}
\centerline{University of Chicago, Chicago, IL 60637}
\vskip .2in

\noindent
The causal boundary of string propagation -- defined as the
hypersurface in loop space bordering the timelike(spacelike)
domains in which two successive measurements of
the string field do(do not) interfere with one another --
is argued to be
  $$0=\int d\sigma\bigl(\delta X(\sigma)\bigr)^2=
	\sum_{\ell=-\infty}^\infty
	\delta x^\mu_{-\ell}\delta x_{\mu\;\ell}\ .$$
Some possible consequences are discussed.

\Date{4/93}

\def\np{Nucl. Phys. }
\def\pl{Phys. Lett. }
\def\pr{Phys. Rev. }

\def\journal#1&#2(#3){\unskip, \sl #1\ \bf #2 \rm(19#3) }
\def\andjournal#1&#2(#3){\sl #1~\bf #2 \rm (19#3) }
\def\nextline{\hfil\break}
\def\ie{{\it i.e.}}
\def\eg{{\it e.g.}}

\def\inbar{\,\vrule height1.5ex width.4pt depth0pt}
\def\IB{\relax{\rm I\kern-.18em B}}
\def\IC{\relax\hbox{$\inbar\kern-.3em{\rm C}$}}
\def\ID{\relax{\rm I\kern-.18em D}}
\def\IE{\relax{\rm I\kern-.18em E}}
\def\IF{\relax{\rm I\kern-.18em F}}
\def\IG{\relax\hbox{$\inbar\kern-.3em{\rm G}$}}
\def\IH{\relax{\rm I\kern-.18em H}}
\def\II{\relax{\rm I\kern-.18em I}}
\def\IK{\relax{\rm I\kern-.18em K}}
\def\IL{\relax{\rm I\kern-.18em L}}
\def\IM{\relax{\rm I\kern-.18em M}}
\def\IN{\relax{\rm I\kern-.18em N}}
\def\IO{\relax\hbox{$\inbar\kern-.3em{\rm O}$}}
\def\IP{\relax{\rm I\kern-.18em P}}
\def\IQ{\relax\hbox{$\inbar\kern-.3em{\rm Q}$}}
\def\IR{\relax{\rm I\kern-.18em R}}
\font\cmss=cmss10 \font\cmsss=cmss10 at 7pt
\def\IZ{\relax\ifmmode\mathchoice
{\hbox{\cmss Z\kern-.4em Z}}{\hbox{\cmss Z\kern-.4em Z}}
{\lower.9pt\hbox{\cmsss Z\kern-.4em Z}}
{\lower1.2pt\hbox{\cmsss Z\kern-.4em Z}}\else{\cmss Z\kern-.4em Z}\fi}
\def\IGa{\relax\hbox{${\rm I}\kern-.18em\Gamma$}}
\def\IPi{\relax\hbox{${\rm I}\kern-.18em\Pi$}}
\def\ITh{\relax\hbox{$\inbar\kern-.3em\Theta$}}
\def\IOm{\relax\hbox{$\inbar\kern-3.00pt\Omega$}}

\newdimen\xraise\newcount\nraise
\def\xpoint{\hbox{\vrule height .45pt width .45pt}}
\def\udiag#1{\vcenter{\hbox{\hskip.05pt\nraise=0\xraise=0pt
\loop\ifnum\nraise<#1\hskip-.05pt\raise\xraise\xpoint
\advance\nraise by 1\advance\xraise by .4pt\repeat}}}
\def\ddiag#1{\vcenter{\hbox{\hskip.05pt\nraise=0\xraise=0pt
\loop\ifnum\nraise<#1\hskip-.05pt\raise\xraise\xpoint
\advance\nraise by 1\advance\xraise by -.4pt\repeat}}}
\def\bdiamond#1#2#3#4{\raise1pt\hbox{$\scriptstyle#2$}
\,\vcenter{\vbox{\baselineskip12pt
\lineskip1pt\lineskiplimit0pt\hbox{\hskip10pt$\scriptstyle#3$}
\hbox{$\udiag{30}\ddiag{30}$}\vskip-1pt\hbox{$\ddiag{30}\udiag{30}$}
\hbox{\hskip10pt$\scriptstyle#1$}}}\,\raise1pt\hbox{$\scriptstyle#4$}}
\def\p {\partial}
\def\frac#1#2{{#1\over#2}}
\def\coeff#1#2{{\textstyle{#1\over #2}}}

\def\bra#1{\left\langle #1\right|}
\def\ket#1{\left| #1\right\rangle}

\def \sinh{{\rm sinh}}
\def \cosh{{\rm cosh}}

\def\exp{{\rm exp}}

\catcode`\@=11
\def\slash#1{\mathord{\mathpalette\c@ncel{#1}}}
\overfullrule=0pt
\def\steepslash{\c@ncel}
\def\frac#1#2{{#1\over #2}}

\def\CC{{\cal C}}

\def\II{{\cal I}}

\def\K{{\bf K}}

\def\vareps{\varepsilon}
\def\del{\delta}

\catcode`\@=12
%
\def\ppl{{p^+}}
\def\pmi{{p^-}}
\def\pperp{{\vec p}}
\def\xpl{{x^+}}
\def\xmi{{x^-}}
\def\xmiz{{x^-_0}}
\def\xperp{{\vec x}}

\def\Xpl{{X^+}}
\def\Xmi{{X^-}}
\def\Xperp{{\vec X}}

\def\A{{\bf A}}
\def\aa{{\bf a}}
\def\rgyr{R_{\rm gyr}}
\newsec{Introduction}
In particle field theory, causality is ensured by the fact that
a) the commutator of two free fields vanishes in the spacelike region,
and b) the interactions are local.  Thus, at least in perturbation
theory, no two measurements of the field at spacelike separated
points can interfere with one another.  Surprisingly enough,
the analogous question in string theory has not yet been addressed.
Here we will calculate the string field commutator in flat
spacetime in the light-front gauge.  The commutator will
indeed vanish outside some hypersurface in loop space.
This hypersurface is {\it not} simply related to the light cone
of the underlying point manifold.  We then make an ansatz
concerning the covariant generalization of the result,
and ask whether causality in string field theory may be quite different
than in particle field theory.

\subsec{Light-Front Gauge}

As a warm-up exercise, let us review the
computation of the particle-field commutator.
We will employ light-front quantization, since in the string
case one has a known operator formalism only in light-front gauge.
A scalar field is decomposed as
\eqn\ptfield{
  \phi(x^+,x^-,\vec x)=
	\int_{-\infty}^\infty \frac{d\pperp}{(2\pi)^{d-2}}
	\int_0^\infty\frac{d\ppl}{2\ppl}
	\biggl(\aa_{\ppl,\pperp}\
	e^{[i(\ppl\xmi+\pmi\xpl+\pperp\cdot\xperp)]}
	\ +\  h.c.\biggr)\ ,
}
with $\pmi=(\pperp^{\;2}+m^2)/2\ppl$, and the particle creation/annihilation
operators $\aa_{\ppl,\pperp}$ satisfying
\eqn\accr{
  [\aa_{\ppl,\pperp},\aa_{\ppl',\pperp\,'}^\dagger]=\ppl
	\ \cdot 2\pi\delta_{\ppl,\ppl'}\cdot
	(2\pi)^{d-2}\delta_{\pperp,\pperp\,'}\ .
}
One finds the commutator
\eqn\fieldcom{
  \eqalign{
  	\bigl[\phi(x^+,x^-,\vec x),\phi({\xpl}',{\xmi}',\vec x')\bigr]=&
	   \int_0^\infty\frac{d\ppl}{2\ppl}
	   \biggl(\K^0_\ppl(\delta\xpl,\delta\xmi,\delta\xperp)-
	   \K^0_\ppl(-\delta\xpl,-\delta\xmi,-\delta\xperp)\biggr)\cr
	\K_\ppl^0=\left(\frac{\ppl}{2\pi\delta\xpl}\right)^{(d-2)/2}&
	   \exp\ i\biggl[-\ppl\del\xmi+\frac{m^2}{2\ppl}\del\xpl
	   +\frac\ppl{2\del\xpl}(\del\xperp)^{\;2}\biggr]\cr}
}
The convergence properties of the $\ppl$ integral depend crucially on the
sign of $(\del x)^2=2\del\xpl\del\xmi-\del\xperp^{\;2}$.
For $(\del x)^2<0$, the integration contour can be rotated
onto the positive imaginary axis for the first term in
the commutator, and onto the negative imaginary axis
for the second; both terms are equal and the difference vanishes.
For $(\del x)^2>0$, the rotation cannot be made so that the
integral converges both at $\ppl\rightarrow0$ and
$\ppl\rightarrow\infty$; the two terms in the commutator do not
cancel and the measurements of the field interfere.

Having completed the exercise, the generalization to string theory
is straightforward.  We work in light-front gauge because
a) there is an obvious operator formalism, and b) the
string field is a gauge invariant (physical) observable.
The light-front gauge string field may be decomposed\ref\kaku{M. Kaku
and K. Kikkawa\journal\pr&D10 (74) 1110;
\andjournal\pr&D10 (74) 1823.  \nextline
See also J.F.L. Hopkinson, R.W.Tucker, and
P.A. Collins\journal\pr&D12 (75) 1653; E. Cremmer and J.-L.
Gervais\journal\np&B90 (75) 410.}\ref\others{P. Goddard, J.
Goldstone, C. Rebbi, and C. Thorn\journal\np&B56 (73) 109;\nextline
S. Mandelstam\journal\np&B64 (73) 205.}
\eqn\strfield{
  \eqalign{
  \Phi(\xpl,\xmiz,\vec x(\sigma))=&
	\int_{-\infty}^\infty \frac{d\pperp}{(2\pi)^{d-2}}
	\int_0^\infty\frac{d\ppl}{2\ppl}
	\prod_{\ell=1}^\infty\sum_{\{\vec n_\ell\}}\cr
	\biggl(\A_{\ppl,\pperp,\{\vec n_\ell\}}\;&
	e^{[i(\ppl\xmiz+\pmi\xpl+\pperp\cdot\xperp_0)]}
	\ {\rm H}_{\vec n_\ell}(\xperp_\ell)
	\quad+\quad h.c.\biggr)\ .\cr}
}
In other words, the string gauge invariance is fixed by the
condition $\Xpl(\sigma)=\xpl$, the other longitudinal
string coordinate being (classically) determined by the reparametrization
constraint
\eqn\constraint{
  \Xmi(\sigma)=\xmiz + \int^\sigma d\tilde\sigma
	\Xperp\,'\cdot\vec P(\tilde\sigma)\ ;
}
the transverse coordinates are expanded
\eqn\transverse{
  \Xperp(\sigma)=\xperp_0+\sum_{\ell=1}^\infty (\xperp_\ell e^{i\ell\sigma}
	+\xperp_\ell^* e^{-i\ell\sigma})\
}
(\ie\ $x_{-\ell}=x_{\ell}^*$; for open strings, $x_{-\ell}=x_\ell=x_\ell^*$).
The string creation/annihilation operators obey the
canonical commutation relations
\eqn\Accr{
  [\A_{\ppl,\pperp,\{\vec n_\ell\}},
	\A_{\ppl',\pperp\,',\{\vec n_\ell'\}}^\dagger]=\ppl\cdot
	\ 2\pi\delta_{\ppl,\ppl'}\cdot(2\pi)^{d-2}\delta_{\pperp,\pperp\,'}
	\cdot\delta_{\{\vec n_\ell\},\{\vec n_\ell'\}}\ ,
}
and the ${\rm H}_n$ are the harmonic oscillator wavefunctions
of the string transverse normal modes.
The mass shell condition is
\eqn\mshell{
  0=2\ppl\pmi-\pperp^{\;2}-
	\sum_{{\vec n}_\ell=0}^\infty \ell\;{\vec n}_\ell-m_0^2\ .
}
Strictly speaking, of course, the ground state mass
$m_0^2<0$ for the  bosonic string and
our integral expressions will be badly behaved.  We will treat
this intercept of the leading Regge trajectory as a free parameter
to be adjusted according to our wishes; if one is a stickler,
one can repeat the calculation in the superstring -- the physics is
unchanged (even there we will want to keep $m_0^2$ small and
positive to separate the poles in the propagator, and tend it to
zero only at the end of the calculation).

Once again it is straightforward to evaluate the field commutator:
\eqn\commie{
  \eqalign{\Bigl[\Phi(\xpl,\xmiz,\vec x_\ell,\vec x^*_\ell)\ ,\ &
	\Phi(\xpl',\xmiz',\xperp\,',{\vec x^{*\prime}_\ell})\Bigr]=
	   \int_0^\infty\frac{d\ppl}{2\ppl}\cr
	&\biggl(\K^0_\ppl(\delta\xpl,\delta\xmiz,\delta\xperp)
	   \prod_{\ell=1}^\infty
	   \K^\ell_\ppl(\del\xpl,\xperp_\ell,\xperp_\ell\,')\quad -\cr
	   &\hskip 2cm \K^0_\ppl(-\delta\xpl,-\delta\xmiz,-\delta\xperp)
	   \prod_{\ell=1}^\infty
	   \K^\ell_\ppl(-\del\xpl,-\xperp_\ell,-\xperp_\ell\,')\biggr)\ .\cr}
}
The $\K^\ell_\ppl$ are harmonic oscillator propagators for
the normal modes
\eqn\SHOprop{
  \eqalign{
  \K^\ell_\ppl(\del\xpl,\xperp_\ell,\xperp_\ell\,')&=\left[
	\frac{i\ell}{\pi\sin[(\ell\del\xpl)/\ppl]}\right]^{(d-2)/2}\cr
	\times\ &\ \exp\left[\frac{-i\ell}{\sin[(\ell\del\xpl)/\ppl]}
	\biggl((|\xperp_\ell|^2+|\xperp_\ell\,'|^2)\cos[(\ell\del\xpl)/\ppl]
	-2{\rm Re}(\xperp_\ell^*\cdot\xperp_\ell\,')\biggr)\right]\ .\cr
}}
Now the question arises, what are the analyticity properties of
the integrand in the complex $\ppl$ plane?  For $\K^0_\ppl$,
the answer is as before; for $\K^\ell_\ppl$ one can continue the
integral to the imaginary $\ppl$ axis as
\eqn\continue{
  \eqalign{
  \K^\ell_\ppl(\del\xpl,\xperp_\ell,\xperp_\ell\,')&\qquad{\buildrel
		\ppl\rightarrow i\vareps\ppl \over
	\longrightarrow}\qquad\left[
	\frac{\vareps\ell}{\pi\sinh[(\ell\del\xpl)/\ppl]}\right]^{(d-2)/2}\cr
	\times\ &\exp\left[\frac{-\vareps\ell}{\sinh[(\ell\del\xpl)/\ppl]}
	\biggl((|\xperp_\ell|^2+|\xperp_\ell\,'|^2)\cosh[(\ell\del\xpl)/\ppl]
	-2{\rm Re}(\xperp_\ell^*\cdot\xperp_\ell\,')\biggr)\right]\ .\cr}
}
The continued contribution to the integrand
coming from the oscillators is regular as $\ppl\rightarrow0$; as
$\ppl\rightarrow\infty$, the exponent becomes
\eqn\propexp{
  -\frac{\vareps\ppl}{\del\xpl}\bigl| \xperp_\ell-\xperp_\ell\,'\bigr|^2
	+O\bigl(\coeff{\ell\del\xpl}{\ppl}\bigr)\ .
}
The sign $\vareps=\pm1$ is chosen according to the sign of $\del\xpl$
so that the exponential is damped.
Combining this result with the zero-mode propagator yields the condition
for convergence of the analytically continued $\ppl$ integral:
\eqn\lcgaugecone{
  2\del\xpl\del\xmiz-\del\xperp_0^{\;2}-
	\sum_{\ell=1}^\infty|\del\xperp_\ell|^2  <  0\ .
}
The total propagator kernel will be continuable, and hence the
two contributions to the commutator will cancel, provided
this condition is satisfied.  Otherwise, if the LHS is positive,
one cannot continue the
integral and the two measurements of the string field will
interfere.

Unfortunately, in the string light-front gauge we find a physical
string field at the expense of knowing where the string is in
the full loop space.  Eq.\constraint\ says that, if we write the string
field in position representation (as we must to look for the
light cone), its transverse momentum is
maximally uncertain and so is $X^-(\sigma)$.  One might try to evade
this by going to minimal uncertainty wavepackets for the string
oscillator modes, but this is only achieved at the expense of smearing
out the string field over the transverse space.  We still lose
control over the spacetime location of the strings.

Nevertheless there is a natural covariant generalization of the
causal boundary, or string light cone
(the vanishing locus of the LHS of \lcgaugecone);
namely
\eqn\covcone{
  {
  0=\sum_{\ell=-\infty}^\infty \delta x^\mu_{-\ell}\delta x_{\mu\;\ell}
	=\int d\sigma\ \del X^2(\sigma)\ .
}}
The restriction of this expression to the light-front gauge
$X^+(\sigma)=\xpl$ is precisely the string light cone determined
above.  It suffers from the problem that, while invariant
under simulataneous reparametrizations of the two loops
$X$, $X'$,  it is not invariant under
independent reparametrizations of them.
However this perhaps should not be expected given the above-mentioned
reparametrization properties of the light-front gauge
causal boundary.  Eq.\covcone\ has the appealing property
that it is Lorentz covariant, in fact it is invariant under the
entire spacetime conformal group, \ie\ Lorentz transformations
generated by $\int d\sigma X^{[\mu}P^{\nu]}$, dilations generated by
$\int d\sigma X\cdot P(\sigma)$, and so on.
In addition it has invariance under rotating the mode amplitudes
among one another, which might be of relevance in the
mythical $\alpha'\rightarrow0$ limit of string theory.

\subsec{Covariant Gauge}

In fact there is an ansatz in the covariant (Feynman-Siegel) gauge
quantization which indeed reproduces the light cone \covcone.
In field theory, the commutator of two fields can be expressed
purely in terms of the Green's function
\eqn\green{
  \bra{0}[\phi(x),\phi(x')]\ket{0} =
	\oint_\CC \frac{dp^0}{2\pi}\int\frac{d\pperp}{(2\pi)^{d-1}}
	\ \frac{i\;e^{ik\cdot x}}{p^2-m^2}
}
where from now on $\xperp$, $\pperp$ will denote $d-1$ component
spatial quantities.  All the standard Green's functions arise
from different choices of contour $\CC$ in the complex $p^0$
plane; the commutator arises from the closed contour that
surrounds both poles in the integrand.  Consider this contour
to be the difference of two contours, one slightly above and
one slightly below the real $p^0$ axis.  Writing the kernel
in the Schwinger representation and doing the $p$ integrals,
we find
\eqn\covptcomm{
  \bra{0}[\phi(x),\phi(x')]\ket{0} =
	\int_0^\infty d\tau \Bigl(\frac{1}{4\pi\tau}\Bigr)^{d/2}
	\biggl(\exp\;i\biggl[\frac{(x-x')^2}{4\tau}-m^2\tau\biggr]
	\quad - \quad h.c.\biggr)\ .
}
In other words we have the same expression as in light-front
quantization with the light-front momentum $\ppl$ replaced by the
inverse of the
Schwinger proper time $\tau$.  The analyticity properties in the
complex $\tau$ plane are thus determined by the sign of $x^2$, and
as before \green\ vanishes outside the light cone.

If we adopt \green\ as a definition of the string field
commutator in covariant quantization, then even though we have
no proper spacetime operator formalism in covariant string field theory
we can evaluate the commutator of two string fields
(for an evaluation of the relevant world sheet path integrals, see
\eg\ \ref\crz{C. Ord\'o\~nez, M. Rubin, and R. Zucchini\journal
\pl&215B (88) 103;\nextline
D. Birmingham and C.G. Torre\journal\pl&205B (88) 289.}
\eqn\covstrcomm{
  \eqalign{\Bigl[\Phi(X(\sigma),B(\sigma),C(\sigma))\ ,\ &
	\Phi(X'(\sigma'),B'(\sigma'),C'(\sigma'))\Bigr]=\cr
	   \int_0^\infty{d\tau}
	\biggl(\K^0_\tau(\del x^\mu_0)&
	   \prod_{\ell=1}^\infty
	   \K^\ell_\tau(x^\mu_\ell,{x^\mu_\ell}')\quad -\quad
	   \K^0_{-\tau}(\del x^\mu_0)
	   \prod_{\ell=1}^\infty
	   \K^\ell_{-\tau}(x^\mu_\ell,{x^\mu_\ell}')\biggr)\ .\cr}
}
with $K^0_\tau(\del x^\mu_0)$ the integrand in \covptcomm\ and
\eqn\covSHOprop{
  \eqalign{
  \K^\ell_\tau(x^\mu_\ell,{x^\mu_\ell}')=\left[
	\frac{i\ell}{\pi\sin(\ell\tau)}\right]^{(d-2)/2}&\cr
	\times\ \exp\biggl[\frac{-i}{\sin(\ell\tau)}
	\biggl(\Bigl(\ell|x_\ell|^2+ & \ell|x_\ell'|^2
	+(c_\ell b_\ell^*+c_\ell^* b_\ell)
	+(c_\ell' b_\ell^{*\prime}+c_\ell^{*\prime} b_\ell')
	\Bigr)\cos(\ell\tau)\cr
	&-2{\rm Re}(\ell x_\ell^*\cdot x_\ell')
	-(c_\ell b_\ell^{*\prime}+c_\ell^* b_\ell')
	-(c_\ell' b_\ell^{*}+c_\ell^{*\prime} b_\ell)
	\biggr)\biggr]\ ,\cr}
}
with $B(\sigma)=\sum b_\ell e^{i\ell\sigma}$
and $C(\sigma)=\sum c_\ell e^{i\ell\sigma}$; we have suppressed
the ghost mode dependence in the arguments of \covSHOprop\
for convenience.
Exactly the same reasoning used above in the analysis
of the light-front gauge commutator leads to the conclusion
that the causal boundary in covariant quantization is
\eqn\fullcovcone{
  \eqalign{
  0=&\del x_0^2 +\sum_{\ell=1}^\infty\bigl(\del x_\ell\cdot \del x_\ell^*
	+\frac1\ell(\del c_\ell\del b_\ell^*
	+\del c_\ell^* \del b_\ell)\bigr)\cr
	=&\int d\sigma\bigl(\del X^2(\sigma) \;+\;
	\del C(\sigma)\frac1\p\del B(\sigma) \bigr)\cr}
}
The restriction of this expression to the bosonic submanifold
of the BRST superspace $\{X,B,C\}$ is indeed \covcone.

\subsec{Consequences}

An interesting issue concerns the relation between the causal
boundary \covcone\ and the propagation of string mass eigenstates.
To pass from the position representation $\Phi(x_0,\{x_\ell\})$
to mass eigenstates $\Phi(x_0,\{n_\ell\})$ one smears the former
against harmonic oscillator wavefunctions $H_{n_\ell}(x_\ell)$.
This entails an uncertainty $\Delta x_\ell\sim 1/\ell$
and the string wanders logarithmically over all spacetime
when all modes are considered.  Clearly some renormalized concept
of string location is needed.  In any case causality for
such smeared objects needs to be formulated.

The surprising thing about \covcone,\fullcovcone\
is that the causal region of
string propagation is not the interior of the light cone
of the underlying point manifold
(defined by the set of pointlike loops).  In other words,
{\it the region of causal contact of a string is NOT
the union of the interiors of the point manifold light cones of the points
along the string}.
Consider the case where one loop is pointlike at the origin
and the other has zero extent in $X^0\equiv t$.  Then Eq.\covcone\ says that
the string fields at these arguments interfere if the spatial sphere
bounded by the radius
of gyration $\rgyr^2=\sum_1^\infty |x_\ell|^2$ of the future string
obeys  $t^2-r^2\ge \rgyr^2$ ($r$ is the radial coordinate of the
center of mass of the second string).  But this restriction
does not forbid much of the string from being outside this
light cone; much of the sphere of radius $\rgyr$ centered
at $r$ may lie outside the point light cone
(since $(r+\rgyr)^2 \ge r^2+\rgyr^2$).

Thus Eq.\covcone\ could prove of tremendous importance.
Since strings interact by joining and splitting, one can imagine
a process by which a piece of a string {\it causally} propagates
{\it outside} the point light cone of a region of the point spacetime;
that part of the string can then split off and influence physics
in what to a pointlike observer is the spacelike region, yet string
causality is obeyed.  Actually this point is confusing.
The split off string does not satisfy the causality condition
with respect to the original string.
A more conservative possibility is that the information propagated
is still carried by the part of the string inside the point light
cone.  This would be the case for instance if the interference of
the two orders of measurement is the same for the set
of all strings which differ only outside one another's
point light cone.  Yet another outcome would be that the split off
string still does not interfere with the original string unless
it is inside the latter's string light cone; then it would still not
be completely in the point spacelike region.
Which of these possibilities is correct is tested when one calculates
the commutator in the presence of a string vertex.

If \covcone\ does reflect the causal structure of string theory,
one must ask if there are black hole solutions to
string theory having a horizon in the {\it string} sense, since
the interior of an ordinary (point spacetime)
black hole might in principle be probed
by doing sensitive interference measurements with strings.
Even if wavepackets cannot communicate across point horizons,
could there be more subtle effects -- for instance, scalar
hair (which has profound influence on the horizon structure)?
The extension of the analysis
to the interacting theory should prove interesting.


\vskip 1in
{\sl Acknowledgements}:
It is a pleasure to thank J. Harvey, D. Kutasov, and
P. Windey for discussions.

\listrefs
\bye